\documentclass[preprintnumbers,twocolumn,prl,amsmath,amssymb]{revtex4}

\usepackage{graphicx}
\usepackage{dcolumn}
\usepackage{bm}

\usepackage{hyperref}

\graphicspath{{Figures/}}

\begin{document}

\title{Perfect preferential orientation of nitrogen-vacancy defects\\
 in a synthetic diamond sample}
\author{M. Lesik$^{1}$}
\author{J.-P. Tetienne$^{1,2}$}
\author{A. Tallaire$^{3}$}
\email{alexandre.tallaire@lspm.cnrs.fr}
\author{J. Achard$^{3}$}
\author{V. Mille$^{3}$}
\author{A. Gicquel$^{3}$}
\author{J.-F. Roch$^{1}$}
\author{V. Jacques$^{1,2}$}
\email{vjacques@ens-cachan.fr}
\affiliation{$^{1}$Laboratoire Aim\'{e} Cotton, CNRS, Universit\'{e} Paris-Sud and Ecole Normale Sup\'erieure de Cachan, 91405 Orsay, France}
\affiliation{$^{2}$ Laboratoire de Photonique Quantique et Mol\'eculaire, Ecole Normale Sup\'erieure de Cachan and CNRS UMR 8537, 94235 Cachan, France}
\affiliation{$^{3}$ Laboratoire des Sciences des Proc\'ed\'es et des Mat\'eriaux, CNRS and Universit\'e Paris 13, 93340 Villetaneuse, France}

\begin{abstract}
We show that the orientation of nitrogen-vacancy (NV) defects in diamond can be efficiently controlled through chemical vapor deposition (CVD) growth on a (111)-oriented diamond substrate. More precisely, we demonstrate that spontaneously generated NV defects are oriented with a $\sim97\%$ probability along the [111] axis, corresponding to the most appealing orientation among the four possible crystallographic axes. Such a nearly perfect preferential orientation is explained by analyzing the diamond growth mechanism on a (111)-oriented substrate and could be extended to other types of defects. This work is a significant step towards the design of optimized diamond samples for quantum information and sensing applications.
\end{abstract}

\maketitle

\begin{figure}[t]
\includegraphics[width = 8.8cm]{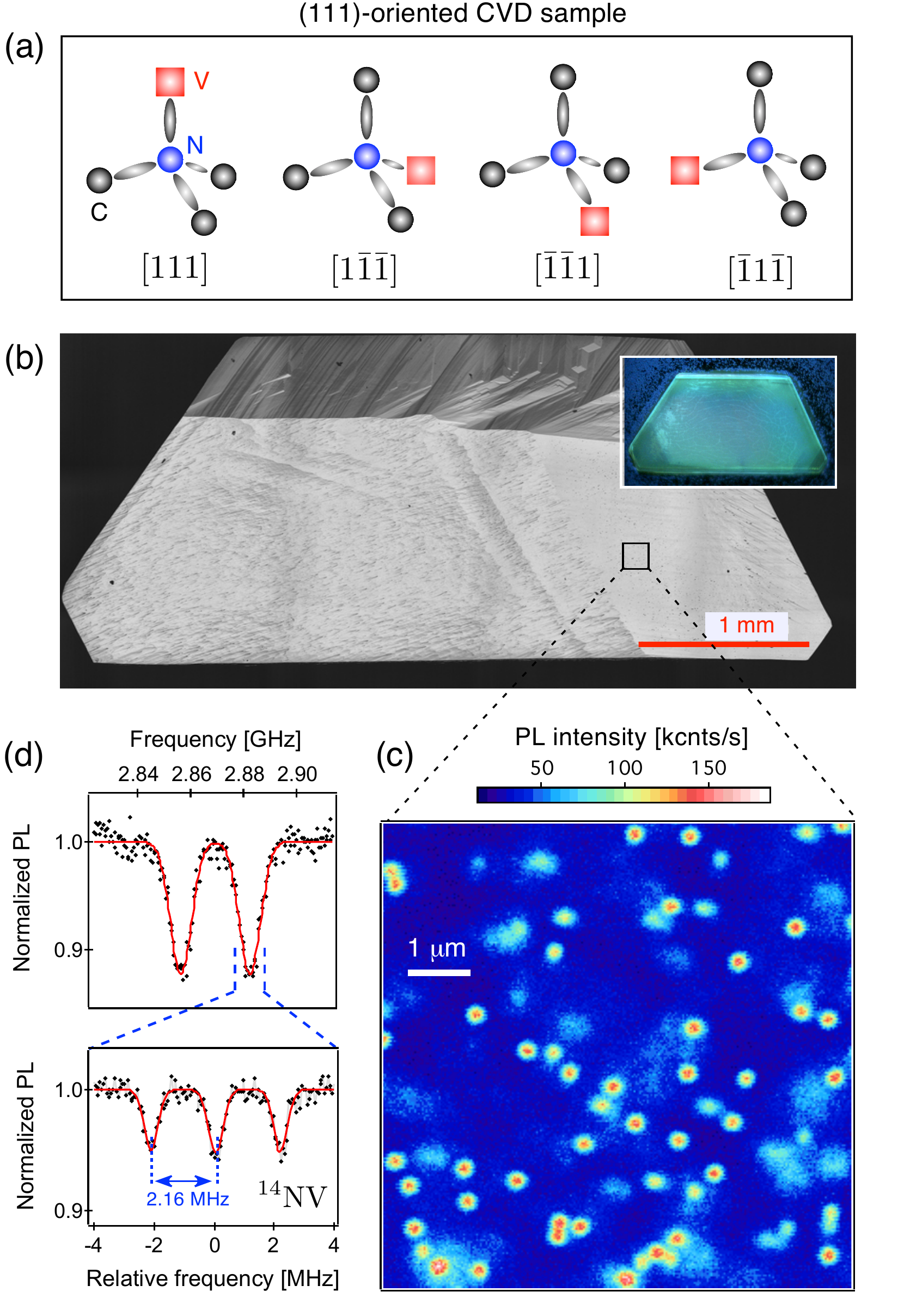}
\caption{(a) Schematic drawing of the four possible orientations of NV defects in a (111)-oriented diamond sample. (b) Microscope image of the studied diamond sample consisting of a 50-$\mu$m-thick CVD film grown on a (111)-oriented substrate. The inset indicates the sample luminescence recorded under ultraviolet light excitation with a DiamondView$^{\rm TM}$ equipment. (c) PL raster scan of the sample recorded with a scanning-confocal microscope under green laser excitation. (d) ESR spectrum recorded from a single NV defect by measuring the PL intensity while sweeping the frequency of a microwave (MW) field. A static magnetic field $B_{0}\sim 5$~G is applied to lift the degeneracy of $m_s=\pm1$ spin sublevels. A high resolution zoom into one of the ESR lines reveals the hyperfine structure associated with the $^{14}$N nucleus of the NV defect.}
\label{Fig1}
\end{figure}

Spins associated with point-like defects in solids are at the heart of a broad range of emerging applications, from quantum information science~\cite{Koehl2011,Dolde2013,Tyryshkin2012}, to the development of highly sensitive quantum sensors~\cite{Taylor2008,Gary2013,Kucsko2013}. In that context, the negatively-charged nitrogen vacancy (NV) color center in diamond has attracted considerable interest over the past years because its electron spin state can be optically detected~\cite{Gruber_Science1997} and exhibits a long coherence time, even under ambient conditions~\cite{Gopi2009}. This defect consists of a substitutional nitrogen atom (N) combined with a vacancy (V) in one of the adjacent lattice sites of the diamond crystal [Fig.~\ref{Fig1}(a)]. Owing to $\mathcal{C}_{3v}$ symmetry, the NV defect can occur with four different orientations in the diamond matrix, along [$111$], [$1\bar{1}\bar{1}$], [$\bar{1}\bar{1}1$], or [$\bar{1}1\bar{1}$] crystallographic axes [Fig.~\ref{Fig1}(a)]. In most diamond samples, these orientations are occupied with equal probabilities. Such a statistical orientation is an important limitation for various applications including the development of hybrid quantum systems where superconducting qubits are coupled to ensembles of NV defects~\cite{Kubo2011,Zhu2011}, high sensitivity magnetometry~\cite{Acosta2009,LeSage2013,Dumeige2013}, and efficient coupling of NV defects to photonic waveguides or microcavities~\cite{Faraon2011,Riedrich2011,Loncar2013}. \\
\indent It was recently shown that {\it partial} preferential orientation of NV defects can be obtained in diamond samples grown by chemical vapor deposition (CVD) in a step-flow mode either on (110)- or (100)-oriented substrates~\cite{Edmonds2012,Pham2012}. More precisely, NV defects were found with a high probability along two directions $\{[111], [\bar{1}\bar{1}1]\}$ rather than the four possible crystallographic axes. Here we report on a nearly perfect preferential orientation of NV defects in a diamond sample grown by CVD on a (111)-oriented substrate. Combining electron spin resonance (ESR) measurements and polarization-dependent photoluminescence (PL) intensity analysis over a set of $\approx 200$ single NV defects, we show that NV defects are oriented along the [111] axis with a $\sim97\%$ probability. Importantly, we show that preferentially-oriented NV defects preserve a long spin coherence time under ambient conditions. This tailored orientation along the [111] axis is ideal for quantum information and sensing applications, since it maximizes the collection efficiency of NV defect emission in bulk single-crystal samples and provide a well-defined geometry with an electronic spin pointing in a direction normal to the diamond sample surface.\\
\indent The key point of this study is the use of a high-purity diamond sample grown by CVD on a (111)-oriented substrate. Plasma-assisted CVD gives unprecedented control over the diamond growth environment, strong flexibility and leads to extremely high purity films with outstanding characteristics~\cite{Tallaire2006,Gopi2009}. However, CVD diamond growth is usually limited to conventional (100) orientations since epilayers on (111)-oriented substrates are plagued by the formation of penetration twins and defects~\cite{Yan1999,Kasu2003}. This issue stems from the fact that nucleation of a new island on a large (111)-oriented surface can occur with equal probability in a normal or a twinned configuration~\cite{Butler2008}. Suppression of these defective sectors can be experimentally achieved if the growth rate in the $[111]$ direction is much higher than that of the twinned plane. In that case penetration twins are rapidly overgrown by their parent face and cannot develop~\cite{Wild1994}. Based on this approach, high-quality diamond films with thickness as large as 100 $\mu$m can be grown on a (111) surface~\cite{Tallaire2013}. The key parameter is to ensure that growth is inhibited in the $[100]$ direction with respect to the $[111]$ direction. The diamond sample shown in Fig.~\ref{Fig1}(b) was grown under such conditions~\cite{Sup}. Its surface is mostly untwined and smooth, even though differences in roughness can be observed across the sample [Fig.~\ref{Fig1}(b)]. The region in the bottom right corner being smoother than the others, it was chosen for this study. The measured root mean square (rms) roughness was around $150$ nm over a $50 \times 50Ê$ $\mu$m$^2$ area.
 It should be however noted that the crystalline quality of such CVD films remains fairly lower than that of conventional (100) CVD films due to the presence of stress and extended defects that lead to intense blue fluorescence under UV light excitation [see inset in Fig.~\ref{Fig1}(b)]. Although no particular shift of the diamond Raman line was evidenced within our set-up accuracy, the width of the line was as high as $2.2$~cm$^{-1}$, when typical values below $1.7$~cm$^{-1}$ are usually obtained on a (100) orientation. 

\begin{figure*}[t]
\includegraphics[width=18cm]{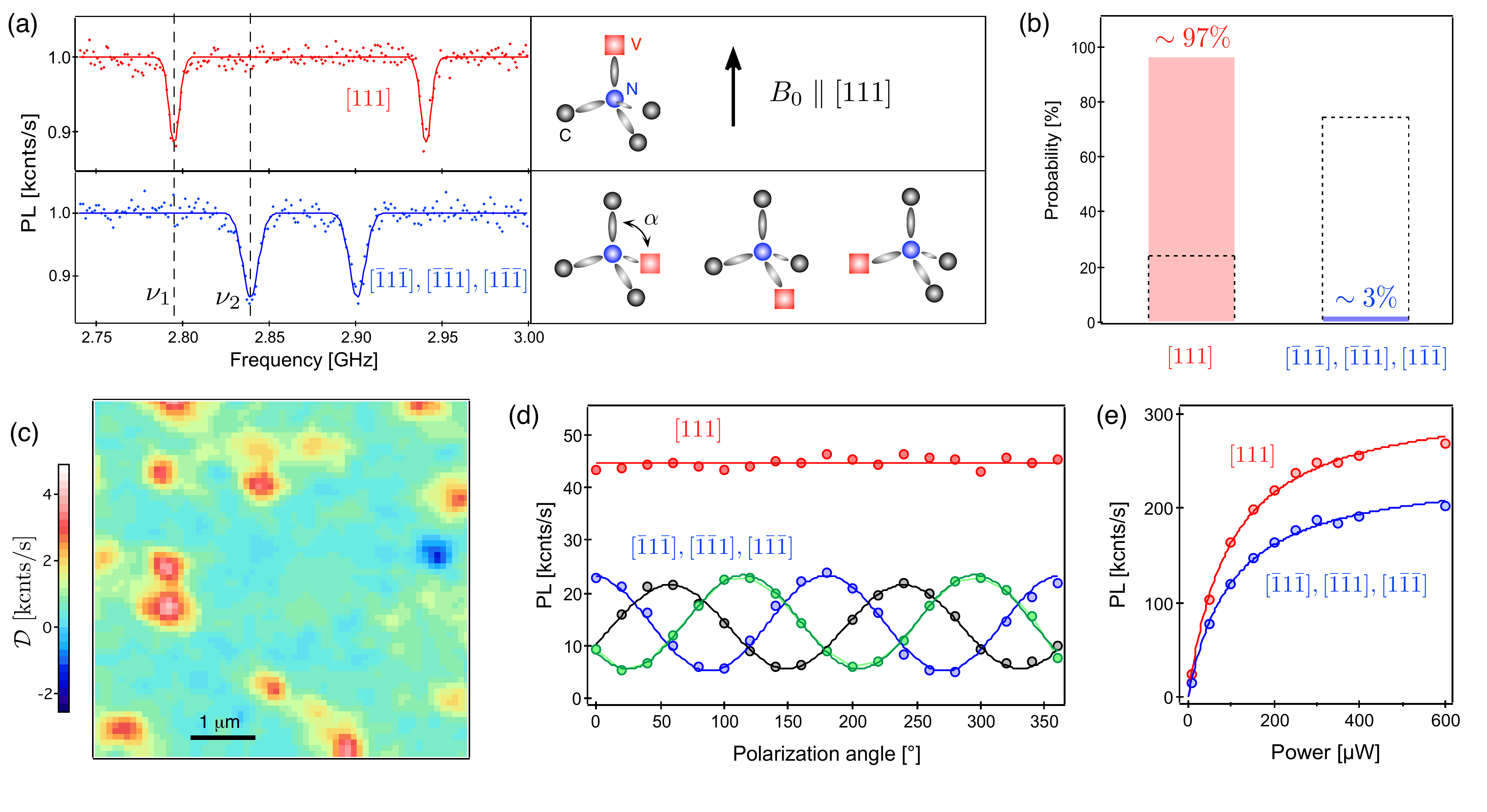}
\caption{(a) Orientation-dependent ESR spectra recorded from single NV defects while applying a static magnetic field $B_{0}=26$~G along the [111] crystal axis. (b) Distribution of NV defect orientations extracted from ESR measurements and polarization-dependent PL intensity for a set of 206 single NV defects. The black dotted bars indicate the expected distribution for randomly oriented NV defects. (c) PL raster scan of the diamond sample recorded by applying consecutively two fixed microwave frequencies $\nu_1$ and $\nu_2$ as shown in (a), while monitoring the difference of NV defect PL intensity $\mathcal{D}$. This signal is positive (red spots) for [111]-oriented NV defects and negative (blue spots) for NV defects with $\{[1\bar{1}\bar{1}],[\bar{1}\bar{1}1],[\bar{1}1\bar{1}]\}$ orientations. (d) Polarization-dependent PL intensity for single NV defects with different crystal orientations. These measurements were performed at weak laser excitation power ($\mathcal{P}=15 \ \mu$W) in order to avoid saturation effects of the optical transition. (e) Collected PL intensity as a function of the laser power for single NV defects having different orientations.}
\label{Fig2}
\end{figure*}

\indent NV defects were optically isolated in this (111)-oriented diamond sample by using a scanning confocal microscope~\cite{Sup}. A typical PL raster scan of the sample is shown in Fig.~\ref{Fig1}(c), showing intense diffraction-limited spots which correspond to single NV defects. This was verified through electron spin resonance (ESR) spectroscopy. The NV defect has a paramagnetic ground state ($S=1$) with a zero-field splitting $D\approx 2.87$~GHz between $m_s=0$ and $m_s=\pm1$ spin sublevels, where  $m_s$ indicates the spin projection along the NV defect quantization axis~\cite{Manson_PRB2006}. Importantly, the NV defect exhibits a spin-dependent PL under optical illumination, which enables ESR measurements by optical means~\cite{Gruber_Science1997}. A typical ESR spectrum recorded from a PL spot is shown in Fig.~\ref{Fig1}(d), revealing the characteristic hyperfine structure associated to the $^{14}$N nuclear spin.\\
\indent The NV defect orientation was first inferred by recording ESR spectra while applying a static magnetic field $B_0$ along the [111] crystal axis [see Fig.~\ref{Fig2}(a)]. The ESR frequencies are then given by $\nu_{\pm}=D \pm g\mu_B |B_{\rm NV}|$, where $g\mu_B\approx 28$ GHz/T and $|B_{\rm NV}|$ is the magnetic field projection along the NV defect quantization axis. For [111]-oriented NV defects $|B_{\rm NV}|=B_0$, while for the three other orientations $|B_{\rm NV}|=B_0|\cos \alpha|$, where $\alpha=109.5^{\circ}$. Typical ESR spectra recorded for a [111]-oriented NV defect and for one of the three other orientations are shown in Fig.~\ref{Fig2}(a), revealing the expected orientation-dependent Zeeman shift of the NV defect electron spin sublevels. In order to estimate the relative population of NV defect orientations, ESR spectra were recorded for a set of $\approx 200$ single NV defects. As depicted in Fig.~\ref{Fig2}(b), we observe [111]-oriented NV defects with a probability of $97 \ \%$. This result indicates an almost ideal preferential orientation of NV defects in (111)-oriented CVD grown diamond samples. For comparison, the probability to find a NV defect along the [111] axis would fall to 25~$\%$ for a diamond sample with random orientation over the four possible crystallographic axes. The high degree of preferential orientation can also be directly evidenced by encoding the NV defect orientation into the PL signal while scanning the diamond sample. For that purpose, we apply consecutively two fixed microwave (MW) frequencies $\nu_1$ and $\nu_2$ at each point of the scan, and measure the difference of NV defect PL intensity $\mathcal{D}={\rm PL}(\nu_{2})-{\rm PL}(\nu_{1})$. As depicted in Fig. 2(a), the MW frequency $\nu_1$ is set on resonance with the electron spin transition linked to [111]-oriented NV defect while $\nu_2$ is resonant for NV defects with $\{[1\bar{1}\bar{1}],[\bar{1}\bar{1}1],[\bar{1}1\bar{1}]\}$ orientations. In that case, the recorded signal $\mathcal{D}$ is positive for [111]-oriented NV defects and negative for all the three other orientations. A PL raster scan of the sample recorded with this dual-MW frequency method is shown in Fig.~\ref{Fig2}(c), confirming the high level of preferential orientation of NV defects in (111)-oriented CVD grown diamond samples.

\begin{figure}[t]
\includegraphics[width=8cm]{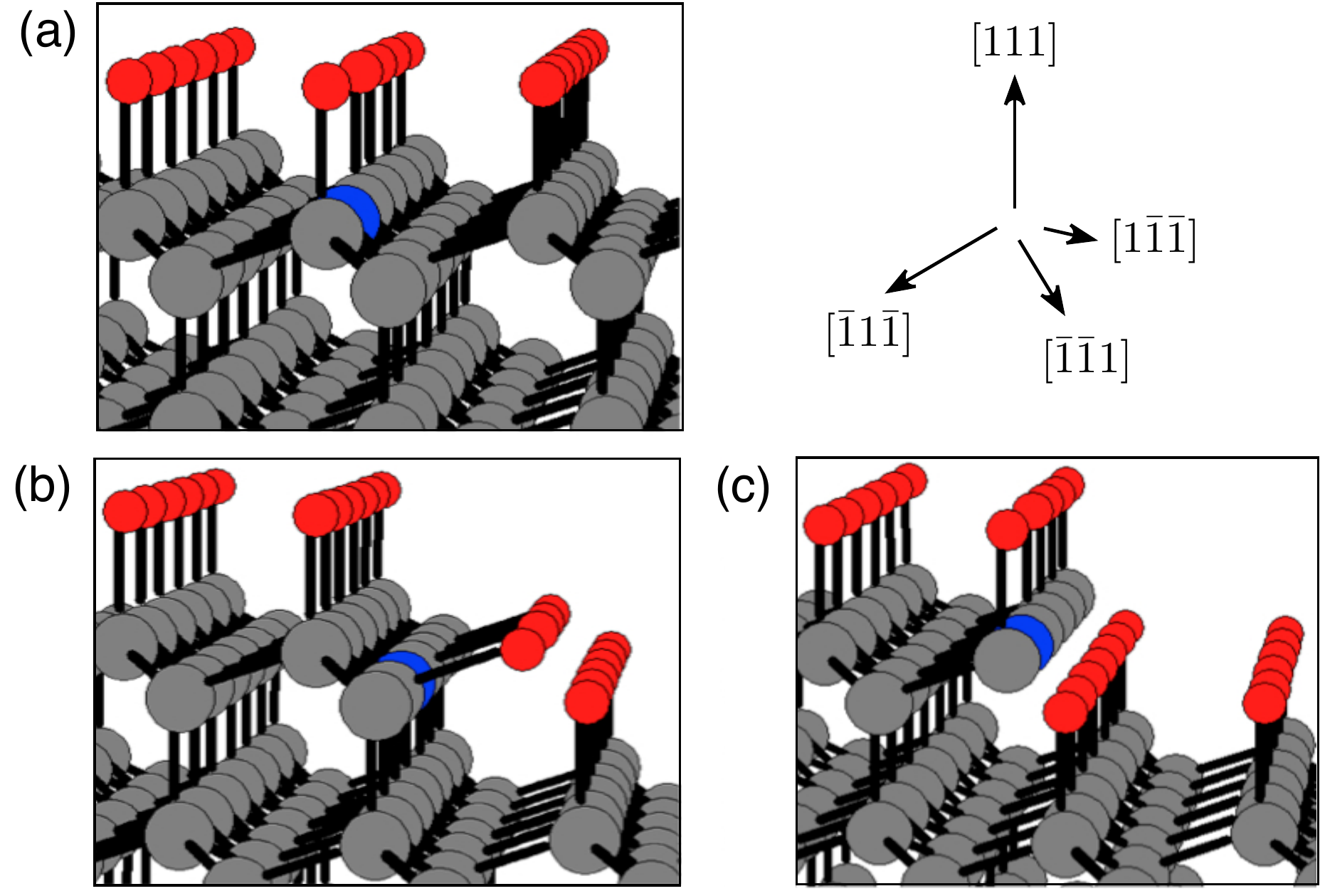}
\caption{Crystal model showing a clean (111)-oriented surface (a) and a surface step requiring the addition of 2 carbons (b) or 1 carbon (c) for the growth to continue. Carbon, hydrogen, and nitrogen atoms are symbolized by grey, red and blue balls, respectively. }
\label{Fig3}
\end{figure}

\begin{figure}[t]
\includegraphics[width=8cm]{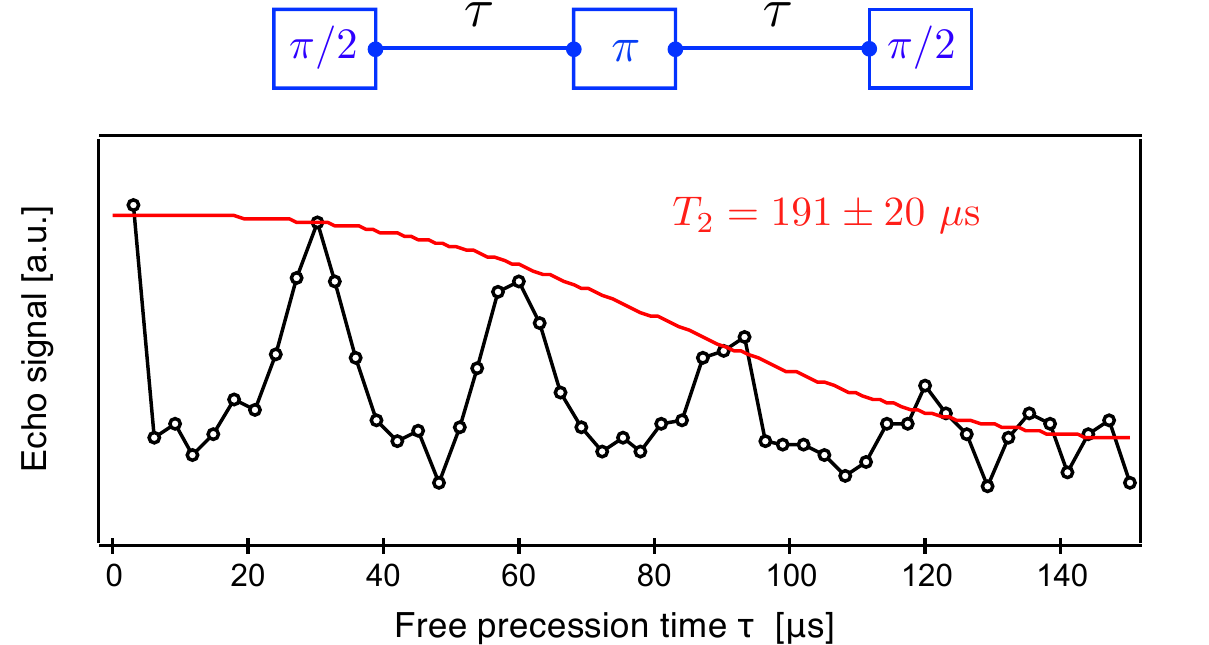}
\caption{Spin echo signal recorded from a single NV defect in a (111)-oriented CVD grown diamond sample, which indicates a coherence time $T_{2}=191\pm 20 \ \mu$s. A static magnetic field $B_0=31$~G is applied along the [111] axis. Similar results were obtained for a set of 10 randomly chosen NV defects. }
\label{Fig4}
\end{figure}

\indent For each single NV defect, the orientation inferred through ESR measurements was further confirmed by recording the PL intensity while rotating the polarization of the excitation laser~\cite{Santori_PRB2007}. Electric dipole transitions are allowed for two orthogonal dipoles lying in the plane perpendicular to the NV axis~\cite{Epstein2005}. The effective excitation rate of the NV defect is then linked to the projection of these dipoles along the linearly-polarized laser field, {\it i.e.} in the diamond sample surface plane. For [111]-oriented NV defects, the two orthogonal dipoles are equally excited in our geometry, and the PL intensity does not depend on the laser polarization angle. Conversely, for NV defects with $\{[\bar{1}\bar{1}1],[1\bar{1}1],[\bar{1}11]\}$ orientations, unbalanced excitation of the two dipoles leads to a polarization-dependent PL intensity, as shown in Fig.~\ref{Fig2}(d). This complementary method enables us to distinguish between all four NV defect orientations in (111)-oriented diamond samples~\cite{Santori_PRB2007}. We finally note that the PL signal from [111]-oriented NV is always higher, even at saturation of the optical transition, since the collection efficiency is also maximized for this particular orientation [Fig.~\ref{Fig2}(e)]. This is a further advantage for quantum information and sensing applications since non-ideal collection efficiency is another limitation to sensitivity improvements.\\
\indent In order to qualitatively explain the observed preferential orientation, we adapt a growth mechanism proposed by Edmonds et al.~\cite{Edmonds2012} for a (110) surface. For that purpose, it is useful to represent a clean, perfectly oriented (111) surface in a CVD environment [Fig.~\ref{Fig3}(a)]. The particularity of such a surface is that it exhibits only one hydrogen-saturated bond pointing towards the [111] direction whereas a (110) surface would have two surface bonds per carbon atom. It was shown that diamond growth on a flat (111) surface first requires the nucleation of a new island by abstraction of hydrogen and bonding of a cluster of 3 carbon atoms~\cite{Battaile1997,Battaile1998}. This process is limiting the growth rate since it takes a relatively long time to occur. On the other hand, once a step is formed at the surface, it flows very quickly by addition of 2 C-atoms [Fig.~\ref{Fig3}(b)] or even 1 C-atom [Fig.~\ref{Fig3}(c)] until full completion of the layer. From this qualitative picture, it appears that if a N atom is incorporated at a step edge during step-flow growth, it is unlikely that it will be directly followed by a vacancy since steps are very attractive for CH$_3$ radicals in the growth environment. In other words it is favourable that the N atom is fully bonded to the surface in a threefold coordination, as illustrated in Fig.~\ref{Fig3}(a). On the other hand, when the surface is overgrown by the next layer, it will occasionally leave a vacancy above the N atom, which is favoured by the fact that it has no available bonding electron to attract another carbon atom. Although this simple picture of the growth process provides insights into the preferential orientation of NV defects, it would be interesting to conduct atomistic modeling in order to calculate precisely the energy of the different configurations of the N atom during (111)-oriented CVD growth~\cite{Atumi2013}. Nevertheless, we note that the reported preferential orientation also confirms that NV defects are grown in as a unit rather than by diffusion of vacancies through the lattice~\cite{Edmonds2012}.\\
\indent Most of the applications involving NV defects rely on the long coherence time of its electron spin state. Since (111)-oriented CVD growth is known to favor incorporation of impurities and defects, the coherence properties of single NV defects were measured and compared to reference values obtained in high purity CVD diamond samples grown on (100)-oriented substrates.  The spin coherence time was measured by applying a spin echo sequence [$\pi /2-\tau-\pi-\tau-\pi/2$], which consists of resonant microwave $\pi/2$ and $\pi$ pulses separated by a variable free evolution duration $\tau$~\cite{Childress_Science2006}. A typical spin-echo measurement recorded from a single NV defect is depicted in Fig.~\ref{Fig4}, showing collapses and revivals of the electron spin coherence induced by Larmor precession of $^{13}$C nuclear spin impurities~\cite{Childress_Science2006}. The envelope of the spin echo signal indicates a coherence time $T_{2}=191\pm 20 \ \mu$s. This value is similar to the one obtained for single NV defect hosted in high purity (100)-oriented diamond samples with the same natural abundance of $^{13}$C (1.1\%). This result shows that incorporation of impurities and defects during (111)-oriented CVD growth does not impair the coherence time of single NV defects. We note that spin coherence time could be further improved either by applying advanced dynamical decoupling protocols~\cite{Lange2010} or by using methane isotopically enriched with $^{12}$C in the CVD reactor~\cite{Gopi2009}.\\
\indent In summary, an almost perfect orientation of NV defects in the [111] direction was achieved by CVD growth on (111)-oriented substrates. This result together with an increased collection efficiency and long spin coherence times paves the way towards high sensitivity magnetometry with ensemble of NV defects~\cite{Acosta2009,Pham2012,LeSage2013,Dumeige2013} and improved coupling with superconducting qubits in hybrid quantum systems~\cite{Kubo2011,Zhu2011}. Furthermore, this orientation is ideal for efficient coupling to photonic waveguide or cavities, since optical dipole are lying in the diamond surface in this geometry~\cite{Faraon2011,Riedrich2011,Loncar2013}. This work, which could be extended to other types of defects in diamond~\cite{Edmonds2012} or silicon carbide materials~\cite{Koehl2011}, therefore provides a significant step towards the design of optimized diamond samples for quantum information and sensing applications.\\

{\it Note:} A related work was simultaneously published in~\cite{Wrachtrup2014}.\\

The authors would like to thank I. N. Kupriyanov and Y. N. Palyanov from the Sobolev Institute of Geology and Mineralogy (Novosibirsk) for providing the high-quality (111) high pressure high temperature (HPHT) substrates used for CVD growth in this study, and P. Maletinsky for fruitful discussions. The research has been partially funded by the European Community's Seventh Framework Programme (FP7/2007-2013) under Grant Agreement n$^{\circ}$611143 (DIADEMS project) and by the French Agence Nationale de la Recherche through the ADVICE project (ANR-2011-BS04-021-03).

\section*{Methods}

 \noindent {\small{\bf CVD diamond growth on (111)-oriented substrates.} (111) plates cleaved and polished from HPHT octahedral-shape synthetic diamond crystals were used as substrates. The crystals, grown by the temperature gradient method using a split-sphere (BARS) apparatus, were provided by the Sobolev Institute of Geology and Mineralogy, Novosibirsk. The plates had a slight offcut along the [111] direction in the range of $2-5^{\circ}$ in order to ease polishing. The plates were thoroughly cleaned in boiling aqua-regia to remove surface contaminations, followed by plasma cleaning. The substrates were then introduced into a home-made plasma reactor chamber operating at high pressure and microwave power (250 mbar, 3500 W). The growth temperature was set to around $1000\pm30^{\circ}$~C. Methane concentration was $1.5$\%. High purity gases were then introduced: H$_2$ was 9N while CH$_4$ was 6N. The growth duration was about 8 hours leading to a thickness of $48 \ \mu$m.}\\
 
\noindent {\small{\bf Experimental setup.} NV defects were optically isolated at room temperature using a scanning confocal microscope. A laser operating at the wavelength $\lambda = 532$~nm is focused onto the diamond sample with a high numerical aperture oil-immersion microscope objective (Olympus, $\times 60$, NA=1.35) mounted on xyz-piezoelectric scanner (MCL, Nano-PDQ375). The red-shifted NV defect PL is collected by the same objective and spectrally filtered from the remaining excitation laser with a dichroic filter and a bandpass filter (Semrock, 697/75 BP). The collected PL is then directed through a 50-$\mu$m-diameter pinhole and focused onto a silicon avalanche photodiode (Perkin-Elmer, SPCM-AQR-14) operating in the single-photon counting regime. For electron spin resonance (ESR) spectroscopy, a microwave excitation is applied through a 20 $\mu$m copper wire directly spanned on the diamond surface. ESR spectra are recorded by sweeping the frequency of the microwave field while monitoring the PL intensity.}

\end{document}